\documentclass[twocolumn,prl,superscriptaddress,showpacs,preprintnumbers,amsmath,amssymb]{revtex4-1}

\usepackage[pdftex]{graphicx}
\usepackage{epstopdf}
\usepackage{graphicx}
\usepackage{dcolumn}
\usepackage{bm}
\usepackage{amssymb}
\usepackage{wasysym}
\usepackage{amsmath} 
\usepackage{amsthm}
\usepackage{bibtopic}
\usepackage{sidecap}

\begin{document}

\def\Ef{$E_{\rm F}$}
\def\Tc{$T_{\rm C}$}
\def\kpara{{\bf k}$_\parallel$}
\def\kperp{{\bf k}$_\perp$}
\def\dirGX{$\overline{\rm \Gamma}-\overline{\rm X}$}
\def\dirGY{$\overline{\rm \Gamma}-\overline{\rm Y}$}
\def\pntG{$\overline{\rm \Gamma}$}
\def\pntX{$\overline{\rm X}$}
\def\invA{\AA$^{-1}$}
\def\G{$\Gamma$} 
\def\Z{$Z$}

\title{Surface Fermi arc connectivity in the type-II Weyl semimetal candidate WTe$_{2}$}

\author{J. S\'anchez-Barriga}
\email[Corresponding author. E-mail address: ] {jaime.sanchez-barriga@helmholtz-berlin.de.}
\affiliation{Helmholtz-Zentrum Berlin f\"ur Materialien und Energie, Albert-Einstein-Str. 15, 12489 Berlin, Germany}
\author{M. G. Vergniory}
\affiliation{Donostia International Physics Center, 20018  Donostia/San Sebastian, Spain}
\affiliation{Department of Applied Physics II, Faculty of Science and
Technology, University of the Basque Country UPV/EHU, Apdo. 644, 48080 Bilbao, Spain}
\author{D. Evtushinsky}
\affiliation{Helmholtz-Zentrum Berlin f\"ur Materialien und Energie, Albert-Einstein-Str. 15, 12489 Berlin, Germany}
\author{I. Aguilera}
\affiliation{Peter Gr{\"{u}}nberg Institute and Institute for Advanced Simulation, Forschungszentrum J{\"{u}}lich and JARA, D-52425 J{\"{u}}lich, Germany}
\author{A. Varykhalov}
\affiliation{Helmholtz-Zentrum Berlin f\"ur Materialien und Energie, Albert-Einstein-Str. 15, 12489 Berlin, Germany}
\author{S. Bl\"{u}gel}
 \affiliation{Peter Gr{\"{u}}nberg Institute and Institute for Advanced Simulation, Forschungszentrum J{\"{u}}lich and JARA, D-52425 J{\"{u}}lich, Germany}
\author{O. Rader}
\affiliation{Helmholtz-Zentrum Berlin f\"ur Materialien und Energie, Albert-Einstein-Str. 15, 12489 Berlin, Germany}

\begin{abstract}

We perform ultrahigh resolution angle-resolved photoemission experiments at a temperature T=0.8 K on the type-II Weyl semimetal candidate WTe$_{2}$. We find a surface Fermi arc connecting the bulk electron and hole pockets on the (001) surface. Our results show that the surface Fermi arc connectivity to the bulk bands is strongly mediated by distinct surface resonances dispersing near the border of the surface-projected bulk band gap. By comparing the experimental results to first-principles calculations we argue that the coupling to these surface resonances, which are topologically trivial, is compatible with the classification of WTe$_{2}$ as a type-II Weyl semimetal hosting topological Fermi arcs. We further support our conclusion by a systematic characterization of the bulk and surface character of the different bands and discuss the similarity of our findings to the case of topological insulators.
\end{abstract}
 

\maketitle

Weyl semimetals (WSMs) have recently attracted a lot of attention due to their unique electronic structure and transport properties \cite{Wan-PRB-2011, Balents-Viewpoint-2011, Balents-Weyl-Multilayer-2011, Bernevig-Weyl-2015, Hosur-transport-Weyl-2013}. They are predicted to host relativistic electrons appearing in the form of spin-polarized topological Fermi arcs \cite{Wan-PRB-2011, Bernevig-Weyl-2015, Hasan-NatComm-Weyl-2015} and to exhibit extremely high mobilities which are crucial for enhancing the efficiency of future electronic devices \cite{Hasan-NatComm-Weyl-2015, Yang-PRB-Weyl-Applications-2011}. These properties offer a rich platform for realization of a huge magnetoresistance \cite{Baibich-PRL-1988, Gruenberg-PRB-1989}, with values exceeding the ones achieved in semiconductors and metals so far \cite{Felser-NatPhys-Transport-Weyl-2015}, paving the way for novel applications of WSMs in future information technologies as well as in spintronics \cite{Yang-PRB-Weyl-Applications-2011, Felser-WeylSpintronics-arXiv-2016}.

In WSMs, the topological Fermi arcs arise as low-energy excitations on the surface \cite{Hasan-NatComm-Weyl-2015, Sun-topological-Fermi-arc-PRB-2015}, and connect the projections of the so-called Weyl points which appear in pairs of opposite chirality \cite{Wan-PRB-2011, Balents-Viewpoint-2011, Bernevig-Weyl-2015}. This is different from Dirac semimetals, where the three-dimensional (3D) Dirac-crossing point between bulk conduction and valence bands arises from a pair of Weyl points that are degenerate in energy as a result of the crystal symmetry \cite{Balents-Dirac-Weyl-PRB-2012, Young-Dirac-Weyl-PRL-2015}. Therefore, to lift the degeneracy of the projected Weyl points in a 3D Dirac semimetal, breaking time-reversal or space inversion symmetry is required \cite{Balents-Dirac-Weyl-PRB-2012, Wang-BreakDirac-PRB-2015}, which in concurrence with strong spin-orbit coupling, results in a WSM phase with nontrivial Fermi arcs that are distinct from surface states in topological insulators (TIs) \cite{Hasan-RMP-2010}. Thus, 3D Dirac semimetals with strong spin-orbit coupling can be viewed as the parent phase of WSMs, which can be classified into type-I and type-II. In type-I WSMs, which respect Lorentz invariance \cite{Wan-PRB-2011, Bernevig-Weyl-2015}, the projection of the 3D conical dispersion of the bulk bands results in a point-like Fermi surface at the Weyl point \cite{Hasan-NatComm-Weyl-2015}. Recently, type-I WSMs have been confirmed experimentally in non-magnetic materials of the TaAs family of crystals, primarily using angle-resolved photoemission spectroscopy (ARPES) \cite{Xu-TypeI-WSMS-Science-2015, Yang-TypeI-WSMS-NatPhys-2015, Lv-TypeI-WSMS-PRX-2015, Felser-NatMat-ARPES-Weyl-2016}. Although the unambiguous identification of nontrivial Fermi arcs in these experiments has proven a challenging task, certain criteria have been given for this purpose \cite{Belopolski-Criteria-PRL-24April2016}. Type-II WSMs, on the other hand, are Lorentz symmetry breaking and become manifest in a highly-tilted Weyl cone in energy-momentum space \cite{Soluyanov-Theory-Prediction-Nature-2015}. As a result, type-II Weyl points emerge at the boundaries between bulk electron and hole pockets, and the Fermi surface is completely different than that of type-I WSMs \cite{Soluyanov-Theory-Prediction-Nature-2015}. This distinction gives rise to a different kind of Weyl fermions that might be responsible for a number of exotic phenomena that are unique for type-II WSMs, such as various magnetotransport anomalies \cite{Udagawa-Anomaly-arXiv-2016}, squeezing and collapse of the Landau levels \cite{Zu-squeezed-Landau-PRL-2016}, a modification of the anomalous Hall conductivity \cite{Zyuzin-Modified-Hall-JETPLett-2016}, or a magnetic breakdown of quantum oscillations due to Klein tunneling \cite{OBrien-Magnetic-breakdown-PRL-2016}. 

Most recently, WTe$_{2}$, known for its large, non-saturating magnetoresistance \cite{Ali-Magnetoresistance-Nature-2014}, has been predicted to be a type-II WSM hosting topological Fermi arcs \cite{Soluyanov-Theory-Prediction-Nature-2015}. The Weyl points in WTe$_{2}$ are located above the Fermi level, making it difficult to resolve them using conventional ARPES. However, on the (001) WTe$_{2}$ surface it is possible to use conventional ARPES to identify topologically nontrivial Fermi arcs, as the arcs should connect electron and hole pockets with opposite Chern number \cite{Soluyanov-Theory-Prediction-Nature-2015}. Following this criterion, only very recently few ARPES studies have aimed at the identification of nontrivial Fermi arcs in WTe$_{2}$ at temperatures above $\sim$7--10 K \cite{Bruno-arxiv-8April2016, Wang-arxiv-14April2016, Belopolski-arXiv-24April2016, Kaminski-arXiv-18April2016, Kaminski-arXiv-1June2016}. However, on the basis of the interpretations of the available experimental data, consensus on the topological character of the observed surface states has not been reached fully. Moreover, triggered by the theoretical prediction of other potential surface states not yet observed experimentally \cite{Soluyanov-Theory-Prediction-Nature-2015}, a complete picture of the bulk or surface character of the different features contributing to the ARPES spectral function is still missing. 

To investigate these issues, here we perform ultrahigh resolution ARPES experiments at a temperature T=0.8 K with the aim of fully resolving the surface Fermi arc connectivity to the bulk electron and hole pockets in WTe$_{2}$. Our main finding is that the connectivity of the surface state is strongly mediated by distinct surface resonances dispersing near the border of the surface-projected bulk band gap, and at the boundaries of the electron and hole pockets. We argue that the coupling to the surface resonances, which are topologically trivial, is compatible with the identification of WTe$_{2}$ as a type-II WSM. We further support our conclusion by first-principle calculations and a systematic characterization of the bulk and surface character of the different bands. Finally, we discuss the analogy of our observations to the case of TIs.

Photoemission experiments were performed with the ARPES $1^3$ endstation at the beamline U112-PGM2b of the synchrotron radiation source BESSY-II. Measurements were taken at a temperature of 0.8 K with a Scienta R4000 analyzer and under ultrahigh vacuum conditions with a base pressure below $1\cdot10^{-10}$ mbar. The (001) WTe$_{2}$ band structure was characterized using different photon energies varying between 20 and 80 eV and linear p-polarization of the incident light. The angular and energy resolutions were set to 0.2$^{\circ}$ and 1 meV, respectively. The sample surface was prepared by cleaving at $\sim$10 K, and was shown to be highly suitable for ARPES experiments with high resolution. It is known that two possible cleavages ($A$ and $B$), most likely related to two types of surface terminations \cite{Kaminski-arXiv-18April2016}, lead to two different band structures in WTe$_{2}$ \cite{Bruno-arxiv-8April2016, Kaminski-arXiv-18April2016}. Type $A$, where no surface states exist between the electron and hole pockets, was not observed in the present experiments. This fact might be related to the cleavage conditions, as we consistently observed only one type of cleavage. Therefore, the present work is focused on the type $B$ electronic structure of WTe$_{2}$, which is the only one relevant to confirm the theoretical prediction \cite{Soluyanov-Theory-Prediction-Nature-2015}. First-principles calculations were performed within the density functional theory (DFT) as implemented in the Vienna ab-initio Simulation Package (VASP) \cite{Kresse1996,Kresse1999}. Projector Augmented Wave (PAW) \cite{Blochl1994} pseudo-potentials were used to represent the valence and core electrons. The exchange correlation was represented within the General Gradient Approximation (GGA) and Perdew-Burke-Ernzerhof for Solids (PBEsol) parametrization \cite{Perdew2008}.

\begin{figure}
\centering
\includegraphics [width=0.5\textwidth]{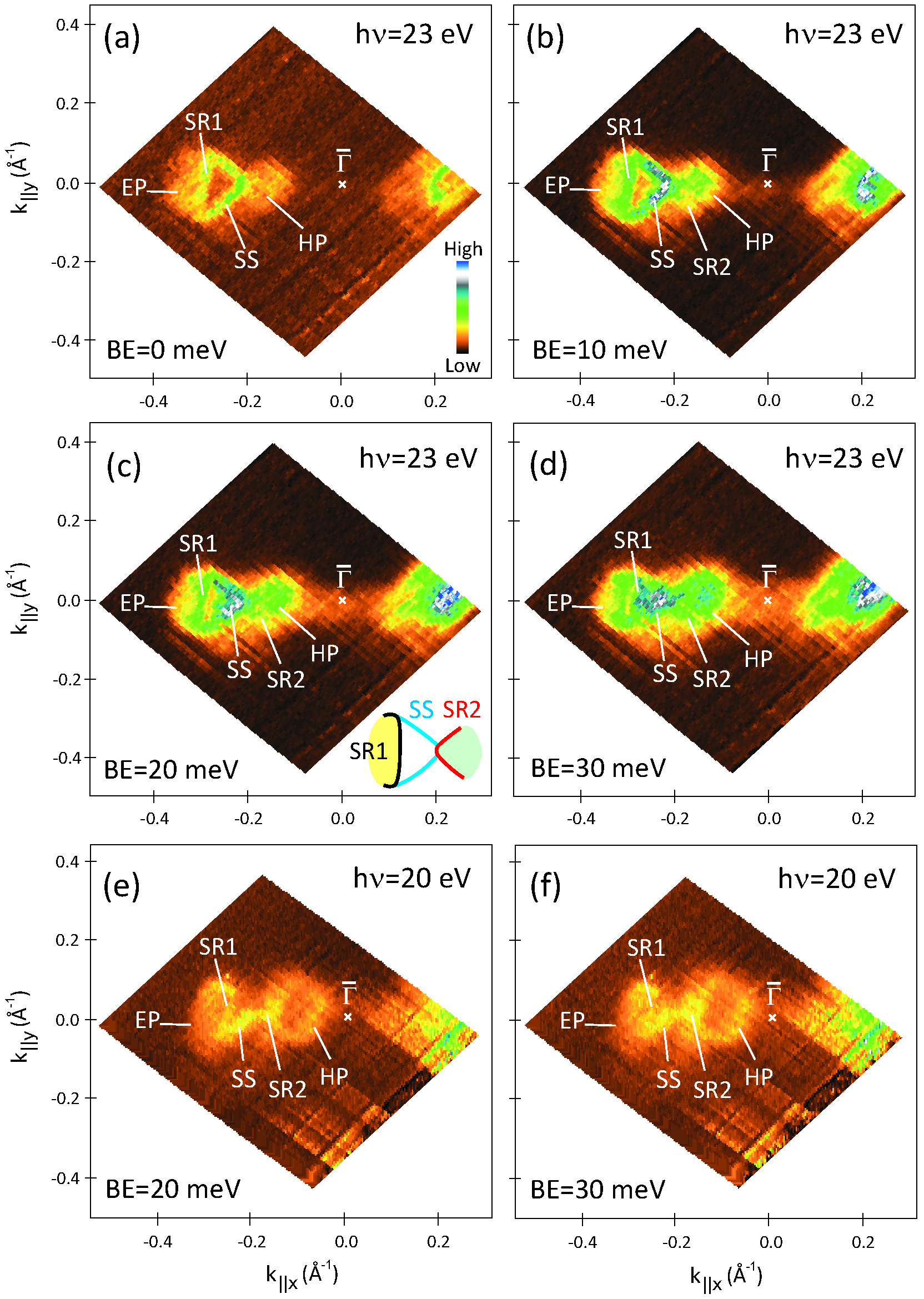}
\caption{High-resolution ARPES maps taken at T=0.8 K with (a)-(d) 23 eV and (e)-(f) 20 eV photons. (a) Fermi surface of WTe$_2$. In (b)-(f), the constant-energy contours are extracted by varying the binding energy (BE) in steps of 10 meV, as indicated in each panel. The different states contributing to the ARPES intensity are labelled accordingly (see text), and their contribution is sketched in the inset of (c). The horizontal momentum axis is parallel to the \dirGX\ high-symmetry direction of the surface Brillouin zone.}
\label{Figure1}
\end{figure}

Figure 1 shows high-resolution ARPES maps taken with 20 eV [Figs. 1(a)-1(d)] and 23 eV [Figs. 1(e)-1(f)] photons. Very clearly, the bulk electron and hole pockets (denoted as EP and HP in Fig. 1, respectively) appear as circular contours which change in size as a function of binding energy (BE). The hole (electron)-like behavior of the bands becomes manifest in the increasing (decreasing) size of the circular contours when moving away from the Fermi level [Fig. 1(a)] by up to 30 meV BE [Figs. 1(b)-1(d)]. Changing the photon energy (h$\nu$) probes the momentum perpendicular to the surface $k_{z}$, and thus the dispersion of the bulk bands \cite{Huefner-Springer-1996}. Due to the large lattice constant of WTe$_2$ ($c\sim$14 \AA) \cite{Brown-CrystWTe2-1966}, a change from h$\nu$=20 eV to 23 eV is equivalent to a variation in $k_{z}$ from 2.95 \invA\ to 3.08 \invA. If we compare the constant-energy surfaces extracted at a BE of 20 meV in Figs. 1(c) and 1(e), the size of the electron and hole pockets also changes as a function of $k_{z}$ due to their 3D character. However, a noticeable $k_{z}$ dispersion is not evident in Fig. 1 for any other states contributing to the ARPES intensity, indicating that they contain a high surface contribution. The most prominent feature is the arc-like structure (labelled SS in Fig. 1) connecting the bulk electron and hole pockets, at first glance consistent with the theoretical prediction \cite{Soluyanov-Theory-Prediction-Nature-2015}. In fact, as the two pockets are formed by well-separated bands, this result immediately suggests that we indeed observe the theoretically predicted topological Fermi arc. 

\begin{figure}
\centering
\includegraphics [width=0.45\textwidth]{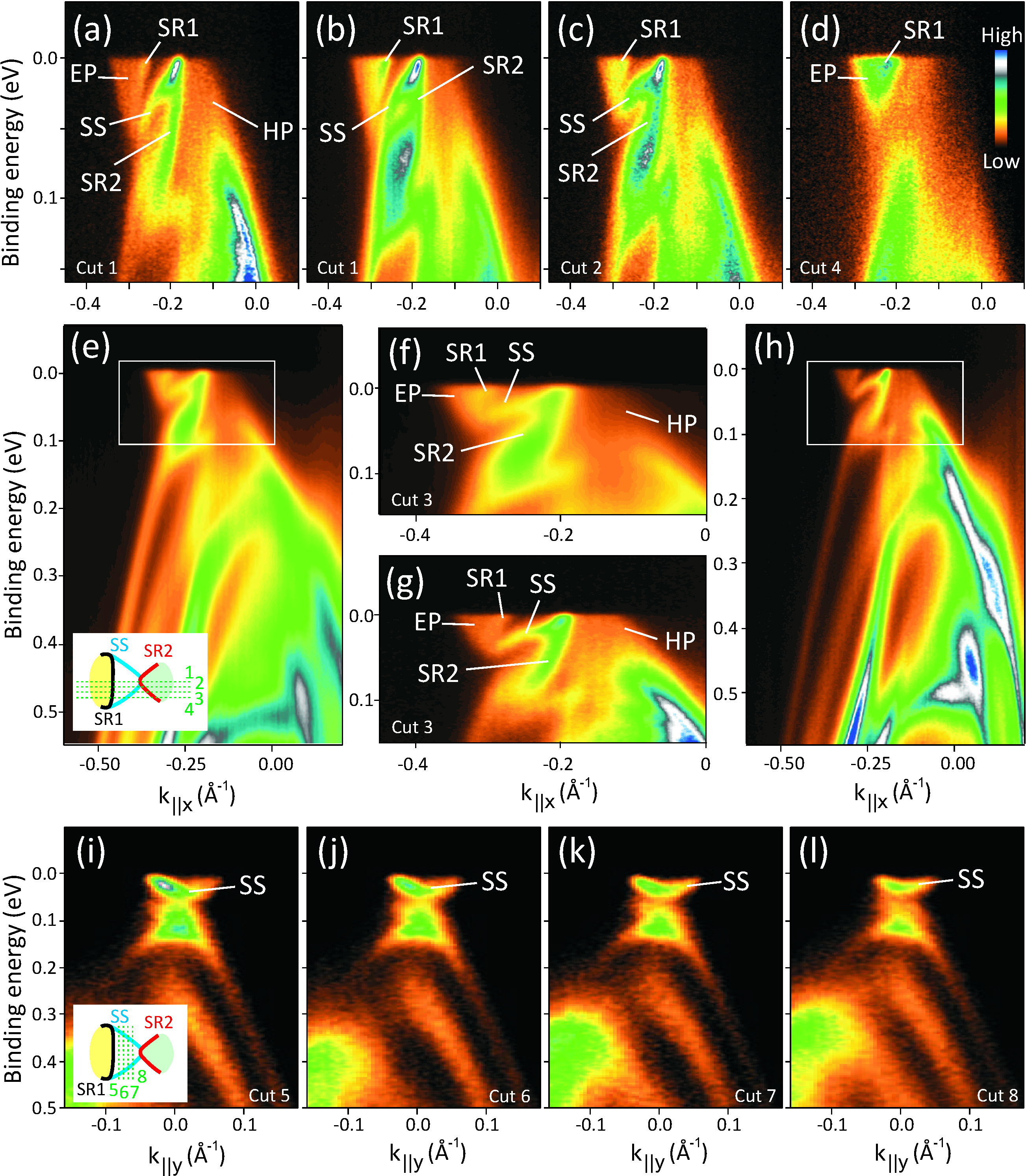}
\caption{Energy-momentum dispersions obtained along momentum cuts that are parallel to the \dirGX\ [(a)-(h)]  and \dirGY\ [(i)-(l)] high-symmetry directions of the SBZ. The different cuts (labelled from 1 to 8) are sketched in the inset of (c) and (i), respectively. Measurements are taken at h$\nu$=25 eV [(a), (g) and (h)], h$\nu$=23 eV [(b)-(f)] and h$\nu$=80 eV [(i)-(l)].}
\label{Figure2}
\end{figure}

\begin{figure}
\centering
\includegraphics [width=0.45\textwidth]{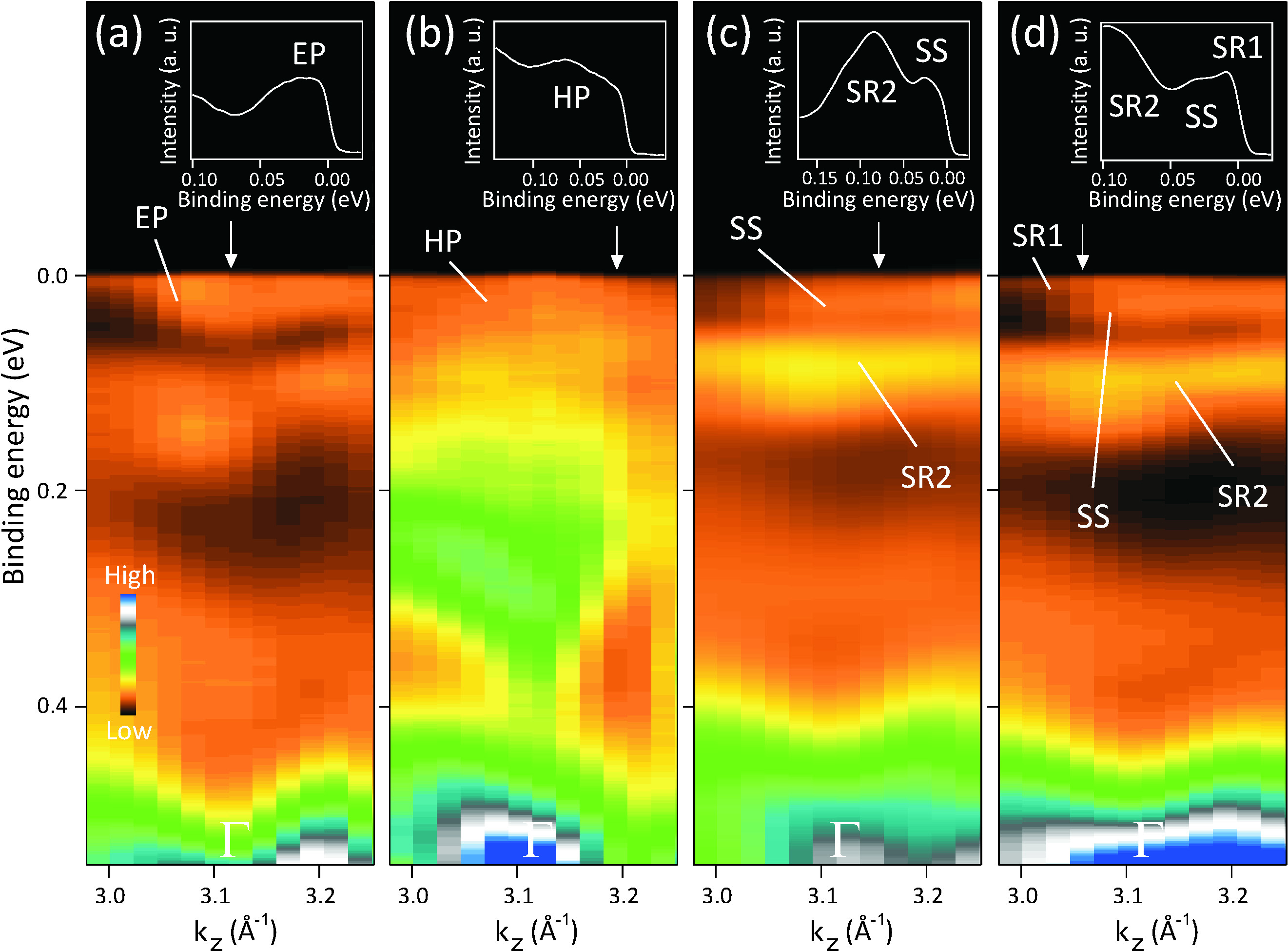}
\caption{$k_{z}$ dispersion of (a) the bulk electron and (b) hole pockets obtained at $k_{x}$=0.36 \invA\ and $k_{x}$=0.18 \invA, respectively. (c)-(d)  Similar results obtained at (c) $k_{x}$=0.29 \invA\ and (d) $k_{x}$=0.32 \invA, containing prominent contributions from SS, SR1 and SR2 states. Measurements are taken around 21 eV photon energy along the \Z-\G-\Z\ direction of the BBZ. Each panel contains an inset with EDCs extracted at the $k_{z}$ values indicated by vertical arrows.}
\label{Figure3}
\end{figure}

A closer look to the constant-energy surfaces of Fig. 1 reveals the existence of other features (labelled as SR1 and SR2) appearing at the borders of the hole and electron pockets. The SR1 feature is most intense at the Fermi level in Fig. 1(a), where it appears as a sharp and straight line connecting the two ends of the arc-like structure at the border of the electron pocket. The SR2 feature, on the other hand, is better seen in Figs. 1(b)-1(d) and connected to the opposite ends of the arc-like structure near the border of the hole pocket. We observe that both SR1 and SR2 states slightly overlap with the bulk continuum, indicating that they are surface-resonance features \cite{McRae-RMP-1979} rather than pure bulk states. The inset of Fig. 1(c) summarizes the contribution from the different states as extracted from the constant-energy maps. Their characteristic dispersion suggests that the connectivity of the arc-like structure to the bulk pockets is directly mediated by the two surface resonances located at the opposite edges of the gap. We point out that these surface resonances are topologically trivial states as they do not connect the electron and hole pockets, in contrast to the Fermi-arc structure observed inside the gap. This finding suggests a complex connectivity between topologically trivial and nontrivial states in the near surface region.

Therefore, to further examine the connectivity between the arc-like surface structure and the bulk electron and hole pockets, in Fig. 2 we show various energy-momentum dispersions obtained along momentum cuts that are parallel to the \dirGX\ [Figs. 2(a)-2(h)] and \dirGY\ [Figs. 2(i)-2(l)] high-symmetry directions of the surface Brillouin zone (SBZ). The different cuts (labelled from 1 to 8) are sketched in the inset of Figs. 2(c) and 2(i), respectively. Measurements are taken at h$\nu$=25 eV [Figs. 2(a), 2(g) and 2(h)], h$\nu$=23 eV [Figs. 2(b)-2(f)] and h$\nu$=80 eV [Figs. 2(i)-2(l)], demonstrating that the sharp and intense surface features can be observed in a wide range of photon energies here corresponding to the sixth [Figs. 2(a)-2(h)] and tenth bulk Brillouin zone (BBZ) [Figs. 2(i)-2(l)]. Note that the $k_{z}$ dispersion of the bulk bands is also apparent by the movement of the top of the hole pocket above the Fermi level at \pntG\ [Figs. 2(a) and 2(b)], or by the filling of intensity inside the electron pocket when comparing Figs. 2(e) and 2(h). 

The characteristic dispersion of the different surface features can be clearly identified in Figs. 2(a) and 2(b) [Cut 1]. In particular, the highest intensity contribution from the SR1 state rims one side of the electron pocket, while the intense SR2 state rims the opposite side of the hole pocket. In addition, the SS feature exhibits a non-linear dispersion connecting the bottom of the electron pocket with the top of the hole pocket through the SR1 and SR2 states, respectively. Such a unique connectivity, which is the manifestation of the surface hybridization, can also be seen in momentum cuts slightly away from the \dirGX\ direction in Fig. 2(c) [Cut 2], Figs. 2(f) and 2(g) [Cut 3] where we zoom-in on the region marked by a white rectangle in Figs. 2(e) and 2(h), respectively. The connectivity eventually disappears when the cut is outside the hole pocket window in Fig. 2(d) [Cut 4]. As expected, the bottom of the SS band lies only slightly below the Fermi level, revealing its electron-like character [see Figs. 2(i)-2(l)].

\begin{figure}
\centering
\includegraphics [width=0.4\textwidth]{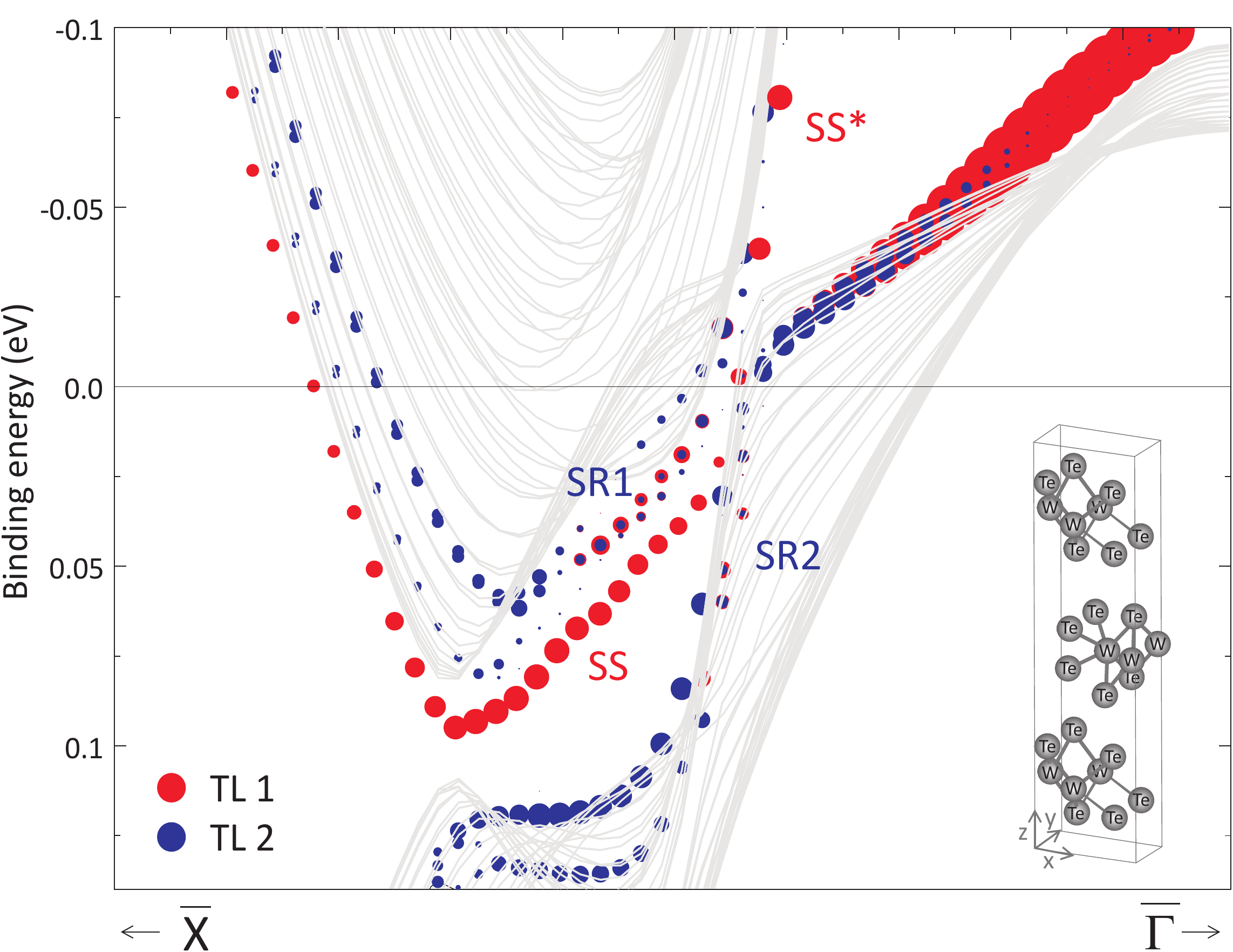}
\caption{Band structure calculation of WTe$_2$ along the \dirGX\ direction of the SBZ. The crystal structure of the topmost surface layers is shown on the right as an inset. The buckled W layer lies between Te layers forming triple layers (TL) stacked with van der Waals bonding. Gray solid lines represent bulk states, blue and red colors indicate the localization of the surface features SS, SR1 and SR2 within the first and second TL from the surface. Size of the color circles represents the orbital weight from Te-$p$ and W-$d$ states in the first and second TL.}
\label{Figure4}
\end{figure}

In Fig. 3, we further confirm the bulk and surface character of the different bands by systematic photon-energy dependent measurements taken around 21 eV  along the \Z-\G-\Z\ direction of the BBZ. The $k_{z}$ dispersions are obtained  along \dirGX\ at various $k_{x}$ points where the intensity contribution from the bulk electron [Fig. 3(a)] and hole pockets [Fig. 3(b)] as well as from the different surface features [Fig. 3(c) and 3(d)] is resolved independently. Each contribution is also emphasized in the corresponding energy-distribution curves (EDCs) shown as an inset. The 3D nature of the electron and hole pockets as well as of other bands at higher BE can be visualized in Figs. 3(a) and 3(b) by a noticeable $k_{z}$ dispersion which is otherwise absent in Figs. 3(c) and 3(d) for SS, SR1 and SR2 states owing to their two-dimensional (2D) character. This scenario is further supported by our first-principles calculations shown in Fig. 4, which reveal qualitative agreement with the experiments. Our calculations show that the surface resonances SR1 and SR2 contain a strong orbital weight from Te-$p$ and W-$d$ states within the topmost surface layers. In particular, both resonant features exhibit much larger localization within the second Te-W-Te triple layer (TL) of the crystal structure as compared to the SS feature which is purely 2D. The qualitative agreement also indicates that the coupling of the SS feature to the electron and hole pockets is governed by $p$-$d$ hybridization through the SR1 and SR2 states, respectively. This result is in principle consistent with the expectation of a highly 2D electronic structure of WTe$_2$ due to its layered structure. However, we point out that the exact contribution from SR1 and SR2 states to magnetotransport is not yet known. 

A similar connectivity through the surface resonances is expected for the second surface state lying above the Fermi level (labelled SS$^*$ in Fig. 4), implying that interruptions of the surface bands around the Weyl points are unlikely. This result is also consistent with the fact that the calculated surface localization of the SS and SS$^*$ features is weakly momentum dependent. Therefore, one can naturally expect this kind of surface connectivity to be a general phenomenon in type-II WSMs. The physical picture underlying the coupling between trivial and nontrivial states revealed by our experiments and calculations in WTe$_2$ has similarities to the one observed previously in TIs \cite{Seibel-2015}. In the latter, distinct surface resonances mediating the coupling between bulk bands and the topological surface state were observed above and below the Fermi level \cite{Seibel-2015, Cacho-PRL-2014, Sanchez-Barriga-PRB-2016}. The difference in the present case is the much narrower dispersion of the arc-like surface feature in momentum space, which in addition remains largely localized within the first TL. This scenario evidences the strong impact of the surface hybridization in WTe$_2$ on the momentum-dependent group velocity of the arc-like feature, in contrast to the case of TIs where the overall coupling between the surface features was found to be weak \cite{Seibel-2015}. Our findings, as well as the details of the surface hybridization discovered here, provide strong circumstantial evidence for WTe$_2$ being a type-II Weyl semimetal hosting topological Fermi arcs. 


{\it Acknowledgments}. This work was supported by the Deutsche Forschungsgemeinschaft (Grant No. SPP 1666) and the Impuls-und Vernetzungsfonds der Helmholtz-Gemeinschaft (Grant No. HRJRG-408). I. A. and S. B. acknowledge financial support from the Virtual Institute for Topological Insulators (VITI) of the Helmholtz Association.

{\it Note added}. During completion of the paper, we became aware of qualitatively similar results posted to arXiv \cite{Bruno-arxiv-8April2016, Wang-arxiv-14April2016, Kaminski-arXiv-18April2016}.








\end{document}